\documentclass[epj,nopacs]{svjour}
\usepackage{graphicx}
\usepackage{latexsym}
\usepackage{amsmath}
\usepackage{amsfonts}
\usepackage{amssymb}
\usepackage{verbatim}
\usepackage[latin1]{inputenc}

\newcommand{\be}{\begin{equation}}
\newcommand{\ee}{\end{equation}}
\newcommand{\bea}{\begin{eqnarray}}
\newcommand{\eea}{\end{eqnarray}}

\newcommand{\ben}{\begin{eqnarray}}
\newcommand{\een}{\end{eqnarray}}

\begin{document}

\title{Confinement and screening in tachyonic matter}
\author{F.A. Brito$^{1}$, M.L.F. Freire$^{2}$, W. Serafim$^{1,3}$, }
\institute{$^{1}$Departamento de F\'\i sica,
Universidade Federal de Campina Grande,
58109-970  Campina Grande, Para\'\i ba,
Brazil\\
$^{2}$Departamento de F\'\i sica, Universidade Estadual da Para\'\i ba, 58109-753 Campina Grande, Para\'\i ba, Brazil
\\
$^{3}$Instituto de F\'\i sica, Universidade Federal de Alagoas, 57072-970 Macei\'o, Alagoas, Brazil}

\date{
%Received: date / Revised version: date
}

\abstract{
In this paper we consider confinement and screening of the electric field. We study the behavior of a static electric field coupled to a dielectric function with the intent of obtaining an electrical confinement similar to what happens with the field of gluons that bind quarks in hadronic matter. For this we use the phenomenon of `anti-screening' in a medium with exotic dielectric. We show that tachyon matter behaves like an exotic way whose
associated dielectric function modifies the Maxwell's equations and affects the fields which results in confining and Coulombian-like potentials in three spatial dimensions. We note that the confining regime coincides with the tachyon condensation, which resembles the effect of confinement due to condensation of magnetic monopoles. The Coulombian-like regime is developed at large distance which is associated with {a screening phase}.
}
\maketitle{\bf Introduction}
\\
\\
The confinement arises naturally in QCD ({\it Quantum Chromodynamics}) where the fields of gluons and quarks appear in a confined state at low energies. Similarly to the electric field, we associate the gluon field charges to ``color charges''. In our studies we show that in general the confining phenomenon can be considered as an effect of a dielectric on the fields and charges with the opposite effect of the screening ``called anti- screening''. {Throughout the paper we shall mention the anti-screening as the way how electric field behaves in a medium with asymptotic behavior that is opposite to that well-known asymptotic behavior in a usual dielectric medium. The hadron matter in its confining phase develops this behavior. The tachyonic field is presented here as a good way to develop both anti-screening (the confinement regime) and screening behavior (the Coulomb-potential behavior at sufficiently large interquark separation).} This seems to be the way in which the hadronic matter lives. In three spatial dimensions this effect presents Coulomb and confinement  potentials to describe the potential between quark pairs. Normally it is used the potential of Cornell \cite{Eichten} $v_{c}(r)=-\frac{a}{r}+br$, where $a$ and $b$ are positive constants, and $r$ is the distance between {\it heavy} quarks. In QED ({\it Quantum Electrodynamics}), the effective electrical charge increases when the distance $r$ between a pair of electron-anti-electron decreases. On the other hand, { in }QCD there is an effect that creates a color charge which decreases as the distance between a pair of quark and anti-quark decreases. Thus, in this sense, the QCD develops effectively opposite QED phenomena. So it is natural to try to understand the QCD phenomena in terms of QED in an ``exotic'' medium such that it creates an anti - screening effect {at} some regime near the QCD scale. In these sense we assume that the electric field is now modified by a ``dielectric function''. Consider $E(r)\equiv\frac{q'(r)}{r^{2}}\equiv\frac{q}{G(r)r^{2}}$ where  $E(r)$ represents the field due to an electric charge $q$ immersed in a polarizable medium playing the role of a ``color" charge coupled to a dielectric function $G(r)$ where $r$ is the distance of the screening (or anti-screening ). The performance of screening in the QED is expressed according to the effective electrical charge i.e.,  $q'(r\gg d)\leq1$ or $G(r\gg d)\geq1$ and $q'(r\ll d)>1$ or $G(r\ll d)<1$ being $d$ the typical size of the screening produced by the polarization of the molecules. On the other hand, in QCD the phenomenon of anti-screening manifests with {color charge} $q'(r\gg R)\gg 1$ or  $G(r\gg R)\ll 1$ and $q'(r<R)\rightarrow1$ or $G(r<R)\rightarrow 1$  and $R$ is the radius of the anti-screening which is the typical size of hadrons. Since we are interested in confinement we will focus our interest in the latter case where the dielectric function $G (r)$ is such that the anti-screening provides the QCD color confinement in a pair of heavy quark. Despite of this, in our present study, a dynamical $G(r)$ comes out in the tachyonic matter, such that for sufficiently large distance the anti-screening breaks down and a screening phase associated with hadronization (or light quarks pair creation) takes place. {Indeed the main propose in the present study is to define a potential model that may develop appropriate potentials for the confining and screening regimes. Whereas the Cornell potential is good to describe heavy quarks, for light quarks a screening regime takes place for sufficiently large interquark separation and another appropriate potential is necessary. In our set up the tachyon matter properly develops the desirable confining and screening behaviors.} Note that we will continue using Abelian gauge fields as in QED, but now $G(r)$ is properly obtained in order to provide both confinement and screening. We will consider only Abelian projections when these fields are embedded in a color dielectric medium \cite{Wilets,Lee}. {Indeed it is already known that results in QCD lattice have been shown that the Abelian part of the string tension accounts  for $92\%$ 
of the confinement part of the static lattice potential. Thus, it suffices to consider only the Abelian (linear part) of the non-Abelian strength field \cite{29,30}}. These facts are useful to study the confinement of quarks and gluons inside the hadrons using a phenomenological effective field theory for QCD \cite{Eichten,Chodos,Bardeen,Friedberg}. There are different ways to confine the electric field in a dielectric medium with different functions $G(r)$ to obtain confinement. In our case the function $G(r)$ is associated with  {\it tachyon condensation} \cite{a-sen}. In general, the behavior of the dielectric function $G$ with respect to $r$ can be governed by a scalar field $\phi(r)$ in which describes the dynamics of the dielectric medium in a {\it tachyon matter}. For this, we use a Lagrangian to describe both the dynamics of gauge and scalar fields associated with tachyon dynamics. {The motivation to use this approach is twofold. First, as we shall see later, this a very natural way to obtain a phenomenological effective field theory for QCD since the aforementioned dielectric function is automatically identified with the tachyon potential. As a consequence, this approach may also address the issues of phenomenological aspects of confinement and screening in string theory. Second, since in this setup the electric confinement is related to tachyon condensation it may bring new insights on confining supersymmetric gauge theories such as the Seiberg-Witten theory \cite{tm76}, which is based on electric-magnetic duality and develops magnetic monopole condensation.}
We shall consider examples of confinement of the electric field in one and three spatial dimensions.

The paper is organized as follows. In Sec.~\ref{sec1} we briefly review the theory of electromagnetism in a dynamical dielectric medium. In Sec.~\ref{sec2} we introduce the tachyon Lagrangian and potentials. We discuss the solutions obtained in one and three spatial dimensions. In the latter case we find analytical potentials under certain conditions and a numeric solution connecting the confining behavior at small distance to screening behavior at large distance. We explore the physics given by the confining and screening potentials. We show how the tachyon condensation is directly related to the confining phase. In Sec.~\ref{conclu} we present our final considerations.
%\\
%\\
\section{\bf Maxwell's equations modified by a dielectric function}
\label{sec1}
%\\
%\\
In this section we apply the theory of electromagnetism in an {\it exotic dielectric medium} to describe the phenomenon of electric confinement. First we write { the Maxwell Lagrangian in the vacuum} without sources
\begin{equation}
\mathcal{L}=-\frac{1}{4}F_{\mu\nu}F^{\mu\nu}.
\end{equation}
As well-known the equations of motion for the electromagnetic field are given by
\begin{equation}
\partial_\mu F^{\mu\nu}=0.
\label{max-ws}
\end{equation}
One should note that even in the absence of sources the equations of motion for the electromagnetic field produce spherically symmetric static vacuum solutions (the Coulomb field due to a point charge) \cite{cvetic} as we shall see below.

For electromagnetic fields immersed in a dielectric medium characterized by the dielectric function $G (\phi)$, where $\phi(r)$ is a scalar field that governs the dynamics of the medium, we have the following Lagrangian
\begin{equation}
\mathcal{L}=-\frac{1}{4}G(\phi)F_{\mu\nu}F^{\mu\nu}.
\end{equation}
The equations of motion (\ref{max-ws}) are now described by
\begin{equation}
\partial_\mu[G(\phi)F^{\mu\nu}]=0,
\end{equation}
being $\mu = 0,1,2,3$. Developing the component $\nu = 0$, we simply have
\begin{equation}
\nabla\cdot[G(\phi)\vec{E}]=0.
\label{eq-3-18}
\end{equation}
We shall be neglecting the magnetic field throughout the paper. This is because the electric field is sufficient for our analysis.

Starting from equation Eq.~(\ref{eq-3-18}) we find the electric field $\vec{E}$ coupled to the dielectric function $G(\phi)$. Now working on spherical coordinates and assuming that $E(r)$ and $\phi(r)$ are  only functions of $r$ and  as a consequence $G(\phi)$ follows the same condition, we have the following form
\begin{equation}\label{eq-6}
\nabla\cdot[G(\phi)\vec{E}]=\frac{1}{r^2}\frac{\partial}{\partial r}(r^{2}G(\phi)E_{r})=0.
\end{equation}
By using this equation we can obtain the vacuum solutions mentioned above. Thus, now we integrate the differential equation to find
\begin{equation}
E_{r}=\frac{\lambda}{r^{2}G(\phi)}.
\end{equation}
Now it is easy to interpret the constant of integration $\lambda={q}/{4\pi\varepsilon_{0}}$ to write the Coulomb electric field modified by a dielectric function $G(\phi)$
\begin{equation}\label{ELET-sol}
E=\frac{q}{4\pi\varepsilon_{0}r^{2}G(\phi)}.
\end{equation}
where $E=|\vec{E}|=E_{r}$. Therefore, we observe that the dielectric function coupled to the electric field $\vec{E}$ changes its magnitude as a function of the radial position $r$.

Let us consider a dielectric function $G(\phi)$ as a function of a dynamical field $\phi$, according to the Lagrangian \cite{Bazeia,Bazeia:2003qr}
\begin{equation}\label{tach-Lag}
\mathcal{L}=-\frac{1}{4}G(\phi)F_{\mu\nu}F^{\mu\nu}+\frac{1}{2}\partial_{\mu}\phi \partial^{\mu}\phi-V(\phi).
\end{equation}
The behavior of the dielectric function $G(\phi)$ will be presented as a consequence of the solutions of the equations of motion obtained by the above Lagrangian.
The equations of motion for the electromagnetic field $A_{\mu}$ and the scalar field $\phi$, are given explicitly by the following differential equations
\begin{equation}
\partial_{\mu}[G(\phi)F^{\mu\nu}]=0,
\end{equation}
\begin{equation}
\partial_{\mu}\partial^{\mu}\phi+\frac{\partial V(\phi)}{\partial\phi}+\frac{1}{4}\frac{\partial G(\phi)}{\partial\phi}F_{\mu\nu}F^{\mu\nu}=0.
\end{equation}
Then the equations of motion for the dielectric medium and electric field in spherical coordinates are
\begin{equation}
\frac{1}{r^{2}}\frac{d}{dr}(r^{2}G(\phi)E)=0,
\end{equation}
\begin{equation}\label{EOM-tach-phi}
\frac{1}{r^{2}}\frac{d}{dr}\left(r^{2}\frac{d\phi}{dr}\right)=\frac{\partial V}{\partial \phi}-\frac{1}{2}E^{2}\frac{\partial G}{\partial \phi}.
\end{equation}
Based on the previous discussion it is easy to show that the solution of the first equation for the electric field is that given in (\ref{ELET-sol}).
%\begin{equation}
%\label{campo12}
%E=\frac{q}{4\pi\epsilon_0 G(\phi)r^{2}}.
%\end{equation}

To find a confining regime {\it everywhere} the dielectric function in (\ref{ELET-sol}) must have the following asymptotic behavior:
\begin{equation}
G(\phi(r))=0\ \ \ \ \ \mbox{as}\ \ \ \ \ \ r\rightarrow\infty,
\end{equation}
\begin{equation}
G(\phi(r))=1\ \ \ \ \ \ \mbox{as}\ \ \ \ \ r\rightarrow 0.
\end{equation}
Particularly, for $G(\phi(\infty))\sim1/r^2$, from Eq.~ (\ref{ELET-sol}) we find $E\equiv const.$ This uniform electric field behavior agrees with confinement.
%Thus we observe that the acoplarmos a function $G(\phi)$ to the electric field $\vec{E}$ in order to obtain the confinement, we note that this function clearly changes the field and the confinement phenomenon can be observed. 
%\\ 
%\\
\section{\bf Tachyon condensation and electric confinement}
\label{sec2}
%\\
%\\
In this section we discuss the relationship between the phenomena of tachyon condensation and the confinement of the electric field. When we speak of ``tachyon'' we refer to particles that are faster than light  and are associated with instabilities. As {\it magnetic monopoles} they have never been observed isolated in nature, although, specially from the superstring point of view, they may always be interacting with other fields or self-interacting at higher orders to form the {\it tachyon condensation} \cite{a-sen,Zwiebach}. In our study we show that the electric confinement  via tachyon condensation can occur in the same way as the confinement of colorful particles, such as quarks and gluons  through {\it condensation of monopoles} \cite{tm76}.
%Throughout this paper we shall adopt the most commonly used  {\it tachyon potential}  of the form
%\begin{equation}\label{exp-tach}
%V=\mbox{c exp}(-\lambda\phi).
%\end{equation}
%\\
%\\
\subsection{\bf Tachyon Lagrangian with electromagnetic fields}
%\\
%\\
For simplicity, we first consider the fields depending only on the spatial component $x$, i.e.
\begin{equation}
\phi=\phi(x),\ \ \ \ \ \ \ \ \ A_{\mu}=A_{\mu}(x).
\end{equation}
Thus, the equations of motion discussed in the previous section are now given by
\begin{equation}
\label{campo5}
\frac{d}{dx}[G(\phi)E]=0,
\end{equation}
\begin{equation}
\label{campo6}
-\frac{d^{2}\phi}{dx^{2}}+\frac{\partial V}{\partial \phi}-\frac{1}{2}\frac{\partial G}{\partial\phi}E^{2}=0,
\end{equation}
where we use the fact that $F^{01}=E$. { Note that Eq.~(\ref{campo5}) is a one-dimensional version of Eq.~(\ref{eq-6})}. Integrating (\ref{campo5}), we have
\begin{equation}
G(\phi)E=q\ \ \ \ \Longrightarrow\ \ \ E=\frac{q}{G(\phi)},
\end{equation}
that substituting into Eq.(\ref{campo6}) we find
\begin{equation}
-\frac{d^{2}\phi}{dx^{2}}+\frac{\partial V(\phi)}{\partial \phi}-\frac{1}{2}\frac{\partial G(\phi)}{\partial\phi}\frac{q^{2}}{G(\phi)^{2}}=0,
\end{equation}
or simply 
\begin{equation}\label{eq-21}
-\phi^{''}+\frac{\partial V(\phi)}{\partial\phi}-\frac{1}{2}\frac{q^{2}}{G(\phi)^2}\frac{\partial G(\phi)}{\partial\phi}=0
\end{equation}
We note that in the above equation we have, in principle, the potential $V(\phi)$ and the function $G(\phi)$. However, we can restrict these choices considering $G(\phi)=V(\phi)$. As we will see below this choice is legitimate when we are working with a Lagrangian that describes the dynamics of tachyons represented by scalar field $\phi$.

As is well-known from string theory, the dynamics of a tachyonic field $T(x)$ coupled with the electric field $E(x)$ is given by \cite{a-sen,Zwiebach}
\begin{eqnarray}
\label{campo18}
e^{-1}{\cal L}&=&-V(T)\sqrt{1-T'^2+F_{01}F^{01}}\nonumber\\
&=&-V(T)\left[1-\frac{1}{2}(T'^2+F_{01}F^{01})+...\right]\nonumber\\
&=&-V(T)+\frac{1}{2}V(T)(T'^2)-\frac{1}{2}V(T)F_{01}F^{01})+...\nonumber\\
&=&-V(\phi)+\frac12\phi'^2-\frac12V(\phi)F_{01}F^{01}+...,
\end{eqnarray}
where $e=\sqrt{|g|}$ in a general spacetime. The power expansion is justified in slow varying tachyon fields, which are suitable to describe tachyon matter \cite{a-sen}.  This derivation remains valid in 3+1 dimensions for $\phi$ dependence with purely radial coordinate $r$ which can be identified with $x$.

In the last equation  (\ref{campo18}) we use the fact that
\begin{equation}
V(T(\phi))=\left(\frac{\partial \phi}{\partial T}\right)^{2}\Rightarrow\frac{1}{2}V(T)(T'^2)=\frac{1}{2}\left(\frac{\partial \phi}{\partial T}\frac{\partial T}{\partial x}\right)^{2}=\frac{1}{2}\phi'^2,
\end{equation}
where $\phi=f(T)$, or $T=f^{-1}(\phi)$.
Note that comparing the equation (\ref{campo18}) with equation (\ref{tach-Lag}) we find the equality $G=V$ is legitimate, so that we can write
\begin{eqnarray}
\label{campo20}
-\phi''+\frac{\partial V}{\partial \phi}-\frac{q^2}{2}\frac{1}{V^2}\frac{\partial V}{\partial \phi}=0, \nonumber \\
-\phi''+\frac{\partial V}{\partial \phi}-\frac{q^2}{2}\left( \frac{-\partial V^{-1}}{\partial \phi}\right)=0, \nonumber \\
\phi''-\frac{\partial V}{\partial \phi}-\frac{q^2}{2} \frac{\partial V^{-1}}{\partial \phi}=0, \nonumber \\
\phi''+\frac{\partial}{\partial \phi}\left[-V-\frac{q^2 V^{-1}}{2}\right]=0, \nonumber \\
\phi''-\frac{\partial \tilde{V}}{\partial \phi}=0.
\end{eqnarray}
{Thus,  Eq.~(\ref{campo20}) is obtained from Eq.~(\ref{eq-21}) using G=V.}
Note that now we are only left with the equation for the scalar field
\begin{equation}
\phi''=\frac{\partial \tilde{V}}{\partial \phi}
\end{equation}
with the potential
\begin{equation}
\tilde{V}=V+\frac{q^2}{2}\frac{1}{V}
\end{equation}
%\\ 
%\\
\subsubsection{\bf Confinement potential for the electric field in a spatial dimension}
%\\
%\\
For a tachyon Lagrangian in Eq.~(\ref{campo18}) expanded polynomially, it is explicitly clear that the dielectric function $G(\phi)$ can be equal to the potential $V(\phi)$. 
Thus, our {\it dielectric function is naturally identified in the context of tachyon theory}. 
Potentials describing the tachyon condensation in string theory are the type that are zero in the vacuum i.e., when $\phi\rightarrow\pm\phi_{vac}$ such that $V(\phi\rightarrow\pm\phi_{vac})=0$.
%$\phi\rightarrow\pm\infty$ such that $V(\phi\rightarrow\pm\infty)=0$.
%Then we apply this formalism to the case of an exponential potential which satisfies the aforementioned condition.

Thus, as an example, we shall adopt the most commonly used  {\it tachyon potential}  of the exponential form
\begin{equation}
V=e^{-\alpha\phi}, \ \ \ \ \mbox{so that}\ \ \ \ \tilde{V}=2\cosh(\alpha\phi)
\end{equation}
for $q^2=2$. The solution of equation (\ref{campo20}) for this kind of potential is given by 
\begin{equation}
\phi(x)=\frac{2}{\alpha}\arcsin[\tan(x)].
\end{equation}
Substituting this solution into the potential $V=e^{-\alpha\phi}$, we find the dielectric function
\begin{equation}
G(x)=V(x)=\frac{1}{(\tan(x)+\mid\sec(x)\mid)^2}
\end{equation}
Therefore, the model reproduces well the behavior of the electric field confinement
\begin{equation}
E(x)=\frac{q}{G(x)}
\end{equation}
Note that in the limit $x\rightarrow\pm\frac{\pi}{2}$ the electric field diverges (``superconfining" system'). On the other hand, in the limit $x\rightarrow0$, the electric field tends to a constant value $E=q$ (``confining'' regime).
%\\ 
%\\
\subsubsection{\bf Confinement and Coulomb potential for the electric field in three dimensions}
%\\
%\\
In three dimensions the extension of Eq.~(\ref{campo6}) (in the absence of magnetic fields) to radial symmetry is direct and given by (\ref{EOM-tach-phi}) that we recast it to the form
\begin{equation}\label{eq-Mov}
-\left[\frac{1}{r^{2}}\frac{\partial}{\partial r}\left( r^{2}\frac{\partial \phi}{\partial r}\right)\right]+\frac{\partial V}{\partial\phi}-\frac{1}{2}\frac{\partial G}{\partial\phi}E^{2}=0   
\end{equation}
Recalling that the solution for the electric field is given by
\begin{equation}
\label{campo7}
E(r)=\frac{q}{4\pi\epsilon_{0}G(\phi)r^{2}}
\end{equation}
and substituting into (\ref{eq-Mov}) we find
\begin{equation}
-\left[\frac{1}{r^{2}}\frac{\partial}{\partial r}\left( r^{2}\frac{\partial \phi}{\partial r}\right)\right]+\frac{\partial V}{\partial\phi}-\frac{1}{2}\frac{\partial G}{\partial\phi}\left[\frac{q}{4\pi\epsilon_{0}G(\phi)r^{2}}\right]^{2}=0   
\end{equation}
Now considering the fact that $G(\phi)$=$V(\phi)$ and
\begin{equation}\label{lambda-q}
\lambda=\frac{q}{4\pi\epsilon_{0}}
\end{equation}
we have
\begin{equation}
-\left[\frac{1}{r^{2}}\frac{\partial}{\partial r}\left( r^{2}\frac{\partial \phi}{\partial r}\right)\right]+\frac{\partial V(\phi)}{\partial\phi}-\frac{\lambda^{2}}{2}\frac{\partial V(\phi)}{\partial\phi}\frac{1}{r^{4}V(\phi)^{2}}=0   
\end{equation}

Assuming again the exponential tachyon potential \\ $V(\phi(r))=e^{-\alpha\phi(r)}$ we have
\begin{equation}
\frac{1}{r^{2}}\frac{\partial}{\partial r}\left( r^{2}\frac{\partial \phi(r)}{\partial r}\right)=\frac{\partial }{\partial{\phi(r)}}\left[e^{-\alpha\phi(r)}-\frac{\lambda^{2}}{2}\frac{e^{\alpha\phi(r)}}{r^{4}}\right]   
\end{equation}
Since $\phi$ depends only on $r$, we can write our equation in terms of ordinary derivatives to  simply have
\begin{equation}\label{our-Eq} 
\frac{d^{2}\phi(r)}{dr^{2}}+\frac{2}{r}\frac{d\phi(r)}{dr}=-\alpha e^{-\alpha\phi(r)}+\frac{\lambda^{2}}{2}\alpha e^{\alpha\phi(r)}\frac{1}{r^{4}}
\end{equation}
For $\alpha\phi(r)$ sufficiently large (which will be easily satisfied for a large rate $r/r_\phi$ --- see below), we can compare our result with the results of Refs.~\cite{R.Dick,Dick} to the confinement of quarks and gluons with $N_c$ colors, i.e.,
\begin{equation}
\label{campo8}
\frac{d^{2}\phi(r)}{dr^{2}}+\frac{2}{r}\frac{d\phi(r)}{dr}=-\frac{g^{2}}{64\pi^{2}f_{\phi}}\left(1-\frac{1}{N_{c}}\right)\exp\left(-\frac{\phi(r)}{f_{\phi}}\right)\frac{1}{r^{4}}
\end{equation}
because the first term on the right side of the equation Eq.(\ref{our-Eq}) becomes negligible.
This now allows us to identify our electric charge $q$ in terms of the color charge of quarks and gluons $g$ as follows
\begin{equation}\label{eq-37}
-\frac{g^{2}}{64\pi^{2}f_{\phi}}\left(1-\frac{1}{N_{c}}\right)=\alpha\frac{\lambda^{2}}{2}
\end{equation}
Recall that from Eq.(\ref{lambda-q}) we have a relation between $\lambda$ and $q$. According to \cite{R.Dick,Dick}, the confining solution is given by 
\begin{equation}
\label{campo10}
\phi(r)=2f_{\phi}\ln\left(\frac{r_{\phi}}{r}\right)
\end{equation}
Manipulating Eq.~(\ref{eq-37}) we have
\begin{equation}
-\frac{1}{f_{\phi}}=\alpha\frac{\lambda^{2}}{2}\frac{64\pi^{2}}{g^{2}}\left(\frac{N_{c}}{N_{c}-1}\right)
\end{equation}
Now identifying $\alpha=-\dfrac{1}{f_{\phi}}$, we find
\begin{equation}
\lambda=\frac{g}{4\pi}\sqrt{\frac{N_{c}-1}{N_{c}}},
\end{equation}
where we have redefined $g\to g/\sqrt{2}$.
Using Eq.(\ref{lambda-q}) we can now write
\begin{equation}\label{eq-45}
q=\epsilon_{0}g\sqrt{\frac{N_{c}-1}{N_{c}}}
\end{equation}
Substituting (\ref{eq-45}) into (\ref{campo7}), with $G(\phi(r))\equiv G(r)$, we have
\begin{equation}
\label{4.2.87}
E(r)=\frac{g}{4\pi G(r)r^2}\sqrt{\frac{N_{c}-1}{N_{c}}}
\end{equation}
being $G(\phi(r))=V(\phi(r))=\exp(-\alpha\phi(r))$, where $\phi(r)$  is the confining solution given in Eq.(\ref{campo10}).
Now using this solution and $\alpha=-\frac{1}{f_\phi}$ we arrive at
\begin{equation}
G(\phi(r))=V(\phi(r))=\exp\left(2\ln\frac{r_\phi}{r}\right)
\end{equation}
or simply
\begin{equation}
\label{campo16}
G(r)=\left(\frac{r_\phi}{r}\right)^2
\end{equation}
Substituting Eq.~(\ref{campo16}) into Eq.~(\ref{4.2.87}), we find the electric field  modified by a ``color dielectric  function'' $G$ developing a `confining medium' which means a confining phase given by
\begin{equation}\label{eqEcte}
E(r)=\frac{g}{4\pi r_{\phi}^2}\sqrt{ 1-\frac{1}{N_c}}
\end{equation}
This solution was first obtained in  \cite{R.Dick,Dick}  in theories with {\it dilatonic} solutions \cite{cvetic}.
The electric field is constant and therefore implies that the lines of force keep together all the time, which features a {flux tube}, a confinement due to an ``anti-dielectric medium'' developed by the hadronic matter that here is represented by `tachyonic matter'. See discussion below.
\\
\\
$\bullet\,${\bf Confining potential}
\\
\\
Integrating (\ref{eqEcte}) we find the confining potential 
\begin{equation}\label{string-tension-s}
v_c(r)=\frac{g}{4\pi r_{\phi}^2}\sqrt{1-\frac{1}{N_c}}\,r+c
\end{equation}
which is a linear confining potential 
\begin{equation}
v_c(r)=\sigma r+c
\end{equation}
where  $c$ is an integration constant and $\sigma$ is the QCD string tension (which in general also depends on $r$) that breaks at some scale of QCD favoring the production of pairs of mesons (hadronization). As we will see, after this scale appears the regime in which a Coulomb-like type potential approaches a constant. Interestingly, our numerical solution to this problem exhibits this behavior. 

\begin{figure}[h!]
	%\centering
		%\includegraphics[scale=0.4]{C:/Users/Wellington/Desktop/Disseração_Wellington/sabado5.eps}
		\includegraphics[scale=0.35]{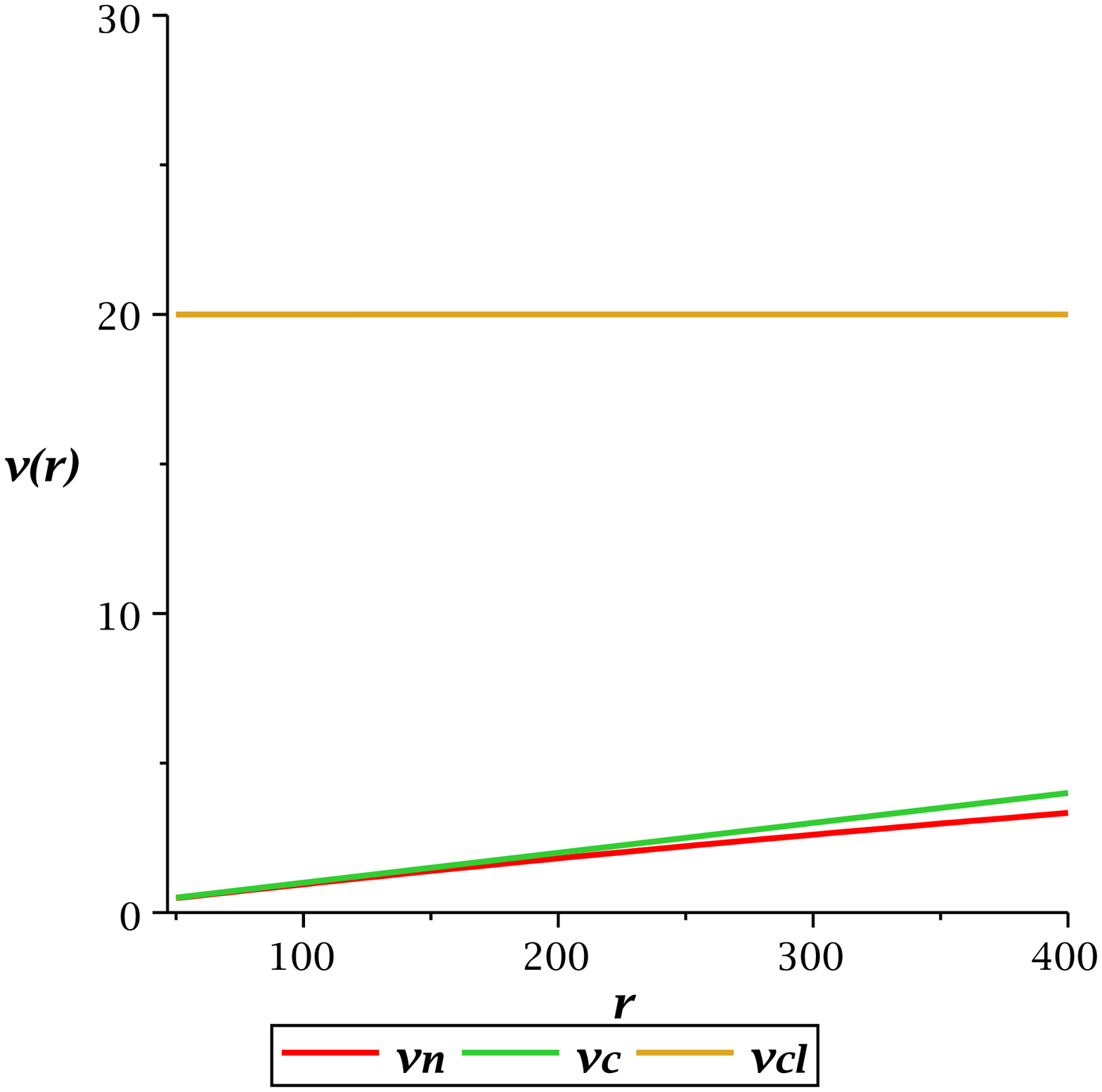}\qquad\qquad\qquad\qquad
		\includegraphics[scale=0.35]{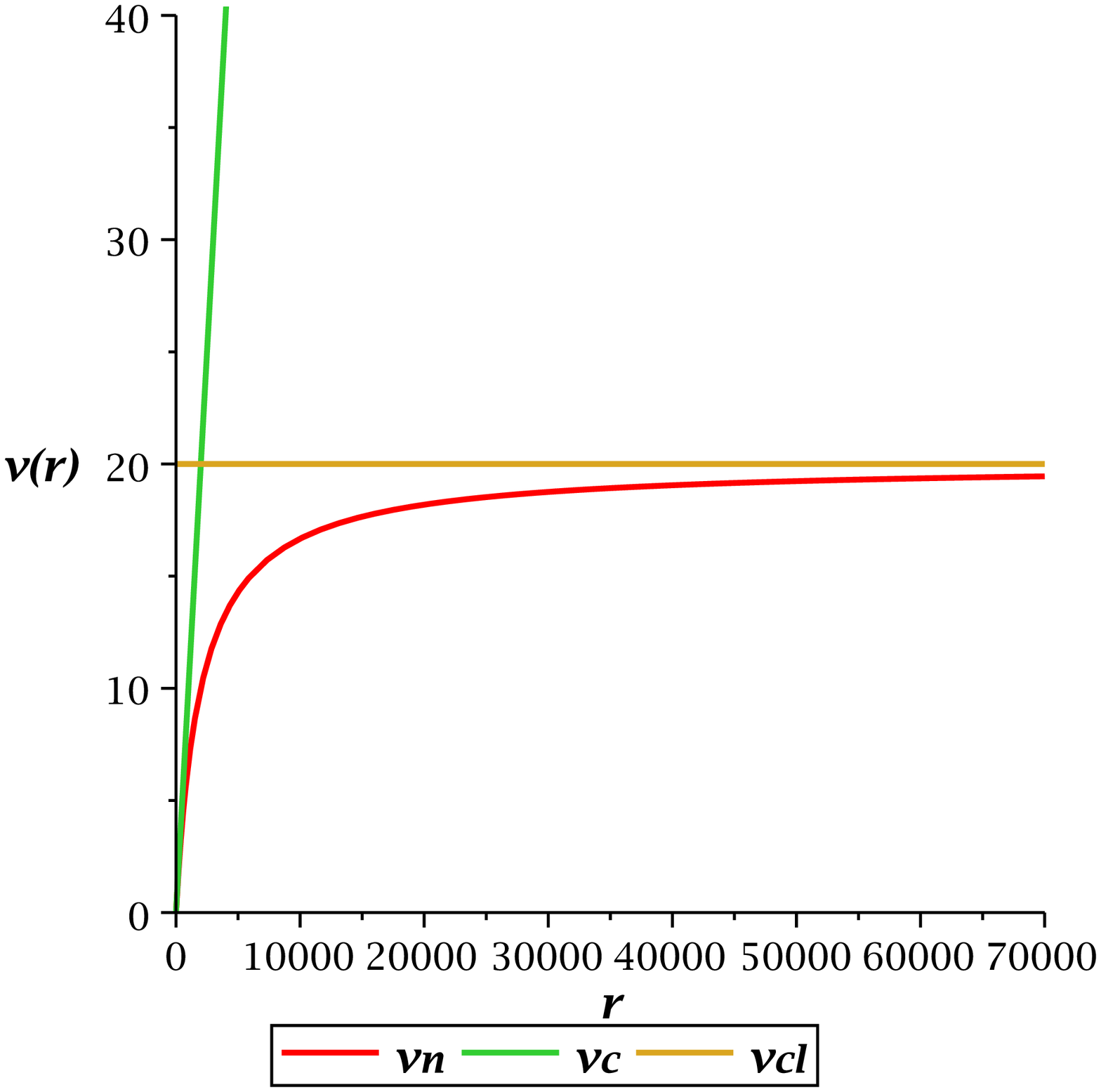}
		\caption{The numerical solution $v_n$ smoothly connects the two regimes in small ($v_c$) and 
large ($v_{cl}$) distances. ({\bf top}) Potentials at small distances:  $v_n$ approaches $v_c$ at a linear confinement phase. ({\bf bottom}) Potentials for large distances:  $v_n$ approaches $v_{cl}$ at a deconfinement (or hadronization) phase.}
	\label{fig1}
\end{figure}

The electric potential energy between two punctual charges is simply given in terms of the potential and the charge 
\begin{equation}
u_c(r)=qv_c(r)
\end{equation}
that is
\begin{equation}\label{eq-51}
u_c(r)=\epsilon_{0}g^{2}\frac{r}{4\pi r^{2}_\phi}\left(1-{\frac{1}{N_c}}\right)
\end{equation}
where for simplicity we have disregarded the integration constant { and in Eq.~(\ref{eq-51}), Eq.~(\ref{eq-45}) has been used}.
So we have a linear confinement type who describes the confinement of quarks and gluons. Note that our treatment is mainly based on an Abelian gauge theory and therefore does not have all the degrees of freedom adequate to describe the colors of gluons. However, as the very expression shows, in the limit $N_c\to\infty$ the electric charge is identical to the color charge. This is the 't Hooft limit (planar limit) where several Feynman diagrams disappear and make the non-Abelian theory simpler. Thus we can understand our approach as an Abelian approach to approximately describe the non-Abelian theory of QCD --- same was considered in \cite{R.Dick,Dick}.
\\
\\
$\bullet$\,{\bf Coulombian potential}
\\
\\
Let us find the Coulomb potential starting from the electric field of Eq.~(\ref{4.2.87}), based on the non-confining solution of Eq.~(\ref{campo8}) for the scalar field \cite{R.Dick,Dick}
\begin{equation}\label{NC-sol}
\phi(r)=2f_{\phi}\ln\left(\frac{r+r_{\phi}}{r}\right)
\end{equation}
we have
\begin{equation}\label{eq-53}
G(\phi(r))=	V(\phi(r))=
%\exp\left(\frac{1}{f_{\phi}}2f_{\phi}\ln\left(\frac{r+r_{\phi}}{r}\right)\right)=
G(r)=\left(\frac{r+r_{\phi}}{r}\right)^{2}
\end{equation}
{ Recall that $f_{\phi} = -1/\alpha$ is used in Eq.~(\ref{eq-53}).}
%that is
%\begin{equation}
%G(r)=\left(\frac{r+r_{\phi}}{r}\right)^{2}
%\end{equation}
Substituting this result into Eq.~(\ref{4.2.87}) we find
\begin{equation}
E(r)=\frac{g}{4\pi{\left({r+r_{\phi}}\right)^{2}}}\sqrt{1-\frac{1}{N_{c}}}
\end{equation}
which is a Coulomb potential regularized at distances $r_\phi$. Integrating the field as a function of $r$ we obtain the `Coulomb' potential as follows
\begin{equation}
v_{cl}(r)=-\frac{g}{4\pi}\sqrt{1-\frac{1}{N_{c}}}\frac{1}{r+r_\phi}+\tilde{c},
\end{equation}
where $\tilde{c}$ is an integration constant. The potential energy is given by
\begin{equation}
u_{cl}(r)=qv_{cl}(r)
\end{equation}
Expressing the charge $q$ in terms of  $g$ as defined above, we found
\begin{equation}\label{eq-57}
u_{cl}(r)=-\epsilon_0 g^2\left(1-\frac{1}{N_{c}}\right)\frac{1}{4\pi(r+r_\phi)}
\end{equation}
where for simplicity we also have here disregarded the integration constant { and in Eq.~(\ref{eq-57}), Eq.~(\ref{eq-45}) has been used}.
The Coulomb-type potential energy for the electric field that mimics the energy of quarks and gluons in the limit of high energies or $r\rightarrow0$. Note, however, that the potential is regular at $r=0$ unlike a usual Coulomb potential. The confining and Coulomb-like potentials are depicted in Fig.~\ref{fig1}. Note that the numerical solution $v_n$ (depicted in red) smoothly connects these two regimes in small and large distances. Here we solve { numerically Eq.~(\ref{our-Eq})} for $\alpha=0.01$ and $\lambda=1$. With this, we use $g/4\pi=1$, $N_c\gg1$, $r_\phi^2=(0.05)^{-1}$ and $\tilde{c}=20$ in potentials $v_c$ and $v_{cl}$. As expected, at small distances the potential approaches the confining solution $v_c$ (consider $c=0$) ---  Fig.~\ref{fig1} - top --- and at large distances the potential approaches a constant given by $v_{cl}$ ( i.e.,  $\tilde{c}=2m_q$ ) --- Fig.~\ref{fig1} - bottom ---. This is the regime where hadronization processes take place via production of light mesons, i.e., for small masses $m_q$.

\begin{figure}[h!]
	%\centering
		%\includegraphics[scale=0.4]{C:/Users/Wellington/Desktop/Disseração_Wellington/sabado5.eps}
		%\includegraphics[scale=0.35]{fig2-potentials-small-r.eps}\qquad\qquad\qquad\qquad
		\includegraphics[scale=0.35]{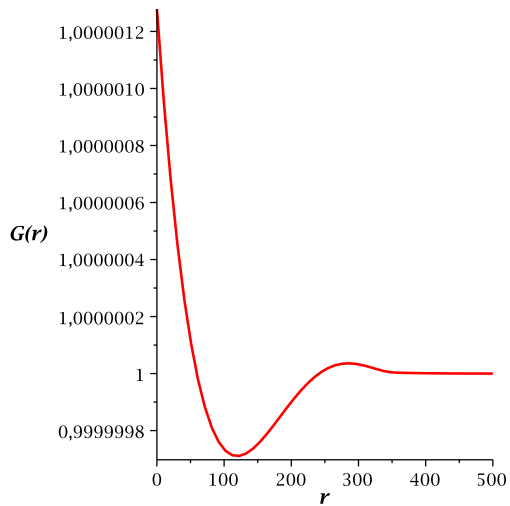}
		\caption{The numerical solution for the dielectric function: $G(r)$ develops a decreasing behavior until  $r\simeq100$ showing a confining (anti-screening) phase and then increases with $r$ developing a deconfining (screening) phase.}
	\label{fig0}
\end{figure}

Finally, we shall comment on the relation between the confinement and the tachyon condensation. We will also comment on the connection in between the confinement and deconfinement phases at small and large distances presented numerically. 

We have found the linear confining by assuming $\alpha\phi$ sufficiently large (i.e., $r\gg r_\phi$)  in Eq.~(\ref {our-Eq}) that leads to Eq.~(\ref{campo8}) which for the tachyon potential
\begin{equation}
V=e^{-\alpha\phi}
\end{equation}
presents the explicit confining solution (\ref{campo10}).  When we substitute this solution into the tachyon potential we find
\begin{equation}\label{cond-1}
V(r)=e^{-\alpha\phi(r)}=\left(\frac{r_\phi}{r}\right)^2
\end{equation}
We note from the equation (\ref{cond-1}) that the tachyon condensation, i.e., $V\to0$, happens when  $r/r_\phi\gg1$. This is automatically satisfied by the hypothesis that $r\gg r_\phi$.

On the other hand, from the numeric analysis, since the dielectric function $G(r)\equiv V(r)$, the Fig.~\ref{fig0} suggests that the confinement/condensation condition is again easily satisfied at $r\simeq100$ since we have chosen $r_\phi^2=(0.05)^{-1}$. 
%As we have seen previously the electric field is written as
%\begin{equation}
%E=\frac{q^2}{4\pi G r^2}
%\end{equation}
%and the confining electric potential is
%\begin{equation}
%v(r)=\sigma r+c
%\end{equation}
We can therefore conclude that the electric confinement is associated with tachyon condensation. 
From the 't Hooft - Mandelstam and Seiberg - Witten \cite{tm76} theories, which deal with the duality in between the phenomena of confinement and superconductivity, we know that the electric confinement can also be associated with the condensation of magnetic monopoles. However, both isolated  tachyons and magnetic monopoles have never been observed.

%seem to have been excluded during the inflationary phase of the Universe.

Now we focus our attention to the deconfining regime. Following a similar analysis and hypothesis using the non-confining solution (\ref{NC-sol}) we found
\begin{equation}
V(r)=\left(\frac{r+r_{\phi}}{r}\right)^{2}\to1
\end{equation}
Again, since $G(r)\equiv V(r)$, the Fig.~\ref{fig0} suggests that the non-confining regime at sufficiently large distance does {\it not} correspond to tachyon condensation. Instead, the full numeric solution depicted in Fig.~\ref{fig1} indeed allows us to identify this regime with a hadronization phase.

Our numerical solution shows that there is a QCD-like string breaking such that the confining is linear only in a certain range, Fig.~\ref{fig1} ({\bf top}), and then the potential becomes constant that is typical of a hadronization phase. 

As a last remark one should note that a large typical hadron size $r_\phi\sim 1/m_q$, which means the limit $m_q\to0$ (light mesons), favors a regime in which a screening phenomenon after the confining phase comes into play, because the QCD string tension given in (\ref{string-tension-s}) becomes small. Then it is expected to be broken easier than for heavy mesons. Thus, for light quarks QCD is never truly confining \cite{Strassler:2001ue}.
%--- see Fig.~\ref{fig0}. 
% {\bf Na fase coulombiana nao ha confinamento electrico nem condensacao taquionica pois neste caso $G(r)\to1$ para $r\to\infty$. Isto eh consistente com fenomeno electrico coulombiano na ausencia de blindagem ou anti-blindagem. A acao taquionica descreve agora uma dinamica DBI-like para ``taquions" e demais campos em todo o espaco.}
%\\
%\\
\section{\bf Conclusions}
\label{conclu}
%\\
%\\
In this study we found a Coulomb-like and confinement potential to the electric field that resembles the ones obtained for the quarks and gluons.  Our Abelian approach can be understood as an approximation of the non - Abelian QCD theory. The confining/deconfining regime of the electric field was obtained considering a dielectric medium whose tachyon dielectric function describes an anti - screening/screening behavior at small and large distance. This medium is regarded as a tachyon matter described in terms of a tachyon potential that vanishes at the minima, that is, in the tachyon condensation. Tachyon condensation was shown to be related to the electric confinement. Thus the tachyon condensation plays the same role as the condensation of monopoles in the electric  confinement phenomenon. The latter corresponds to the duality between the confinement and the phenomenon of the superconductivity.

{\bf Acknowledgments}
\noindent
We would like to thank CNPq, CAPES, PROCAD - CAPES for partial financial support. We also thank Tiago Mariz and Emanuel Cunha for discussions.

\end{document}